\documentclass{emulateapj}
\submitted{{\it Accepted for publication in PASP}}

\usepackage{natbib}
\usepackage{epsfig}

\begin{document}

\title{Updated Colors for Cool Stars in the SDSS}

\author{Andrew A. West}
\affil{Astronomy Department, University of Washington, Box 351580, Seattle, WA 98195-1580}
\email{west@astro.washington.edu}

\author{Lucianne M. Walkowicz}
\affil{Astronomy Department, University of Washington, Box 351580, Seattle, WA 98195-1580}
\email{lucianne@astro.washington.edu}

\author{Suzanne L. Hawley}
\affil{Astronomy Department, University of Washington, Box 351580, Seattle, WA 98195-1580}
\email{slh@astro.washington.edu}

\begin{abstract}
We present updated colors for M and L dwarfs based on photometry from the third data release of the Sloan Digital Sky Survey (SDSS).  These data are improved in quality and number from earlier results.  We also provide updated equations for determining photometric parallaxes from SDSS colors of late-type stars. Walkowicz, Hawley \& West (2004) have recently presented new techniques for studying the magnetic activity of low-mass stars and their method relies on an accurate determination of SDSS color.  We derive new relationships between SDSS colors and other common passbands and present updated formulas from Walkowicz et al. (2004) for determining the level of magnetic activity in M and L dwarfs.
\end{abstract}


\section{Introduction}

The Sloan Digital Sky Survey (SDSS; York et al. 2000; Gunn et
al. 1998; Fukugita et al. 1996; Hogg et al. 2001; Smith et al. 2002;
Pier et al. 2003), as of its third data release (DR3; Abazajian et
al. 2005), has spectroscopically observed approximately 20,000 M and
L dwarfs.  These data have been used to examine magnetic activity
(West et al. 2004; hereafter W04), search for low mass subdwarfs
(W04), and study the properties of white dwarf- M dwarf binary systems
(Raymond et al. 2003) with an unprecedented quantity of data. Hawley
et al. (2002; hereafter H02) first characterized a sample of the M, L and T
dwarfs from the early data release (EDR; Stoughton et al. 2002) of
SDSS.  As a part of their study, they provided average colors for each
spectral type and calculated both spectroscopic and photometric
parallaxes as a function of spectral type and color respectively.
Many current studies are using low mass stars to probe the structure
and composition of the Galaxy (West et al. 2005, in preparation; Juric
et al. 2005, in preparation; Covey et al. 2005, in preparation;
Bochanski et al. 2005, in preparation) and rely on the colors and
photometric parallaxes provided by H02.  However, these colors, and
therefore the photometric parallaxes, were based on the EDR version of
the SDSS photometric pipeline.  Subsequent SDSS data releases have
utilized updated photometric software, resulting in RMS changes of
several hundredths of a magnitude compared to the EDR photometry. The
photometric pipelines for both Data Release 1 (DR1; Abazajian et
al. 2003) and Data Release 2 (DR2; Abazajian et al. 2004) contain
improved software that more rigorously calibrates the data and more accurately
accounts for sky subtraction in the psf photometry.  The SDSS
photometric pipeline was frozen after DR2 ensuring that the DR3
photometry presented in this paper will not change due to SDSS
processing alterations.

The latest SDSS data releases have also dramatically increased the
number of spectroscopically observed stars, allowing us to select a
large sample of high signal-to-noise stars, a luxury not possible with
the smaller EDR dataset.  Our signal-to-noise cuts provide more
accurate spectral types and photometry.  This ability to make quality
cuts on the data and still retain a large sample has substantially
altered some of the colors reported in H02.  In this paper, we report
an updated set of low mass star colors from a high signal-to-noise
subset of the W04 sample, and derive new photometric parallax
relations using the updated photometry of the H02 sample.

In order to aid in the analysis of magnetic activity in W04, a new
technique was designed by Walkowicz, Hawley \& West (2004; hereafter
WHW) for calculating the ratio of the luminosity in H$\alpha$
($L_{\rm{H}\alpha}$) to the bolometric luminosity
($L_{\rm{bol}}$). WHW determined a $\chi$ factor, that when multiplied
by the H$\alpha$ equivalent width (EW), gives the value of
$L_{\rm{H}\alpha}$/$L_{\rm{bol}}$.  The $\chi$ factor varies with
spectral type and WHW used the low mass star colors reported in H02 to
derive $\chi$ as a function of color.  Here we present new equations
for determining $\chi$, based on our updated photometry.

\begin{figure}[h]
\includegraphics[angle=90,scale=.35]{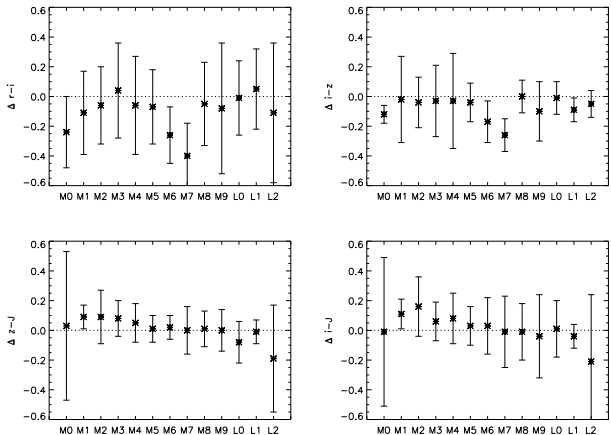}
\caption{Color difference between this study and H02 as function of spectral type for the four colors reported in Table \ref{tbl-1}. Error bars are the spreads reported in H02.  Discrepancies in the $r-i$ and $i-z$ colors are likely due to our large sample size and high signal-to-noise quality cut.  Because the $z-J$ and $i-J$ colors are from the \emph{same} stars as used by H02, the color differences  in these bands indicate changes in the SDSS photometric pipeline.}
\label{fig1}
\end{figure}

\section{Data}

All of the SDSS photometry reported in this study has been reduced using the DR3
version of the SDSS photometric pipeline (Photo 5\_4\_25).  Our main sample 
of low mass stars comes from the W04 spectroscopic sample.  The stars
were selected from the SDSS database based on their $r-i$ and $i-z$ colors 
(see W04 for a detailed description), and have been
spectral-typed using template fitting and molecular band indices as described
in H02.  All spectral types were confirmed by manual inspection.
For this study, we have selected stars from each spectral type bin
with photometric errors less than 0.05 magnitudes in $r$, $i$ and $z$
for M dwarfs and less than 0.05 magnitudes in $i$ and $z$ for
L dwarfs.

We have also downloaded the DR3 photometry for the stars from the H02 study in
order to recalibrate the photometric parallax measurements in the manner of H02.
The same quality cuts on photometric errors that are discussed above have been
applied to these data.  We adopt the 2MASS $J$ magnitudes and the
spectroscopic parallax relations from H02.  The uncertainties in the
absolute $M_J$ magnitudes from that paper are $\sim$ 0.5 magnitudes. 

We use the sample of spectrophotometric stars from WHW to derive new relations
for $\log(\chi)$ as a function of color. We follow the 
exact method of WHW, but utilizing the updated colors from this study.

\begin{deluxetable*}{cccccc}[h]
\tablewidth{0pt}
\tablecaption{Average Color by Spectral Type \label{tbl-1}}
\tablehead{
\colhead{Spectral Type}& 
\colhead{$r-i$}& 
\colhead{$i-z$}& 
\colhead{$z-J$}& 
\colhead{$i-J$}& 
\colhead{$M_{J}$}}
\startdata
M0& 0.67 (0.14)& 0.37 (0.06) & 1.45 ( --- )  &1.92 ( ---) & 6.45\\
M1& 0.88 (0.15) & 0.48 (0.13)& 1.34 (0.09) &1.89 (0.05)&  6.72\\
M2& 1.03 (0.18)& 0.58 (0.18)&  1.45 (0.17)& 2.10 (0.23) & 6.98\\
M3& 1.33 (0.30)& 0.70 (0.31)&  1.46 (0.15)& 2.14 (0.15) & 7.24\\
M4& 1.51 (0.22)&  0.84 (0.22)& 1.57 (0.14)& 2.47 (0.16) & 8.34\\
M5& 1.91 (0.14)& 1.05 (0.09)&  1.66 (0.09)& 2.75 (0.12) & 9.44\\
M6& 2.01 (0.14)& 1.10 (0.07)&  1.76 (0.09)& 3.02 (0.22) &10.18\\
M7& 2.27 (0.20)& 1.26 (0.12)&  1.95 (0.15)& 3.46 (0.24) &10.92\\
M8& 2.77 (0.16)& 1.62 (0.12)&  2.05 (0.13)& 3.71 (0.22) &11.14\\
M9& 2.81 (0.28)& 1.69 (0.07)&  2.23 (0.09)& 4.05 (0.13) &11.43\\
L0& 2.63 (0.27)& 1.84 (0.08)& 2.28 (0.09)& 4.24 (0.10) &11.72\\
L1& 2.61 (0.27)& 1.83 (0.09)& 2.51 ( --- )& 4.41 ( --- )& 12.00\\
L2& 2.39 (0.18)& 1.80 (0.17)& 2.57 (0.11)& 4.43 (0.12)& 12.29\\
\enddata

\tablecomments{The mean and ($\sigma$) of each color are given.
Because these results were calculated using data with small
measurement errors (see Section 2), the $\sigma$ in most cases
represents the intrinsic scatter in the population.  The M0 and L1
bins have only one star each with both $J$-band data and uncertainties
that meet our SDSS data quality cuts; therefore neither have values
for $\sigma$. }

\end{deluxetable*}

\section{Results}
\subsection{Mean Colors}
Using the DR3 colors for the samples discussed above, we calculate updated
$r-i$, $i-z$, $z-J$ and $i-J$ colors for M and early L dwarfs.  
Table \ref{tbl-1}
shows these values sorted according to spectral type.  The mean $r-i$ and $i-z$
colors come from the W04 sample and the $z-J$ and $i-J$ colors 
come from the
updated H02 data.  It is important to note that the parentheses in the table
give the standard deviations of the mean colors and \emph{do not} reflect the
photometric uncertainties (which are at most $\sim$0.07 magnitudes). Instead,
these values reflect the \emph{intrinsic} scatter in the color distribution
at a given spectral type.

To demonstrate the difference between our colors and those presented by H02, we plot the difference in color (this study $-$ H02) as a function of spectral type for the four colors presented in Table \ref{tbl-1} (Figure \ref{fig1}).  The error bars are the spreads given by H02.  Although the mean relations can vary by as much as 0.4 magnitudes, most of the discrepancies fall within the ranges reported by H02.  The $r-i$ and $i-z$ colors that we measure for stars with spectral types M0, M6 and M7 have significant differences from those of H02. These offsets are due to our larger sample size, and our use of only high signal-to-noise SDSS data.  The $z-J$ and $i-J$ data come from the same stars used by H02, but the EDR photometry has been replaced by that of DR3.  Although some of the difference comes from our data quality cut, much of the offset can be attributed to changes in SDSS photometry.

\begin{figure}[h]
\plotone{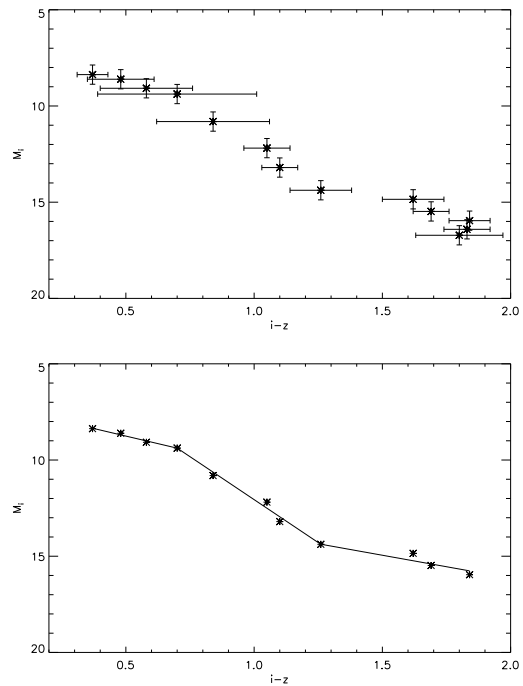}
\caption{Absolute $i$-band magnitude as a function of mean $i-z$ color at 
each spectral type.  The upper panel error bars show the intrinsic 
scatter in the colors and \emph{do not} represent photometric 
uncertainties.  The lower panel gives the same data together with the updated photometric parallax 
relation from Equation 1.}
\label{fig2}
\end{figure}

\begin{figure}
\plotone{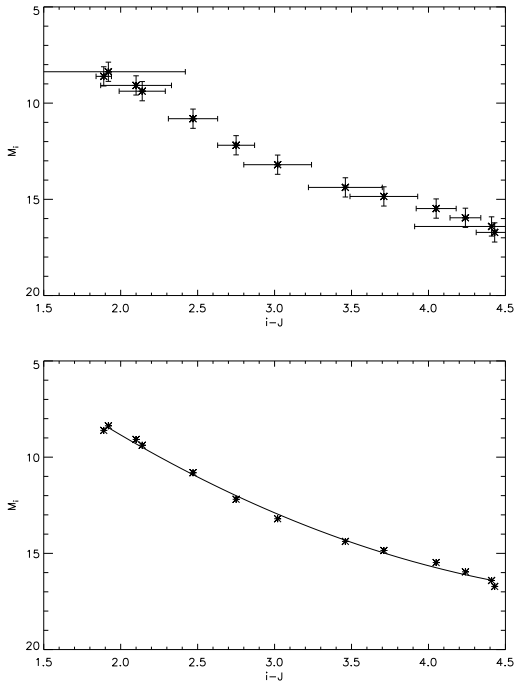}
\caption{Absolute $i$-band magnitude as a function of mean $i-J$ 
color at each spectral type.  The upper panel error bars indicate 
the intrinsic scatter in the colors of low mass stars.  The lower panel
shows the updated photometric parallax relation given in Equation 2.}
\label{fig3}
\end{figure}

Using the mean colors, we derive new photometric parallax relations for the
$i-z$ and $i-J$ colors by plotting the absolute $i$-band magnitude 
as a function of color. We fit the $M_i$ vs. $i-z$ relation with a 3 part linear piece-wise function, which is a better description of this relation than a polynomial fit.  Spectral types later than L0 are not included because of the similar $i-z$ colors for stars cooler than L0.  The $M_i$ vs. $i-J$ relation is best fit by a second order polynomial.   Figure \ref{fig2} and Figure \ref{fig3} show these relations 
for the $i-z$ and $i-J$ colors respectively.  In both figures, the 
error bars on the upper panels
indicate the intrinsic spread in color at each spectral type. The fit to the color-magnitude relation is indicated on the lower panels.   These
fits can be expressed by:

\setcounter{equation}{0}

\begin{eqnarray}
M_{i}(0.37 < i-z < 0.70) = 7.18+3.14(i-z)\;\pm0.08\;\;\;\;\;\nonumber\\
M_{i}(0.70 < i-z < 1.26) = 3.13+8.93(i-z)\;\pm0.25\;\;\;\;\;\\
M_{i}(1.26 < i-z < 1.84) = 11.4+2.39(i-z)\;\pm0.31\;\;\;\;\;\nonumber\\
\nonumber\\
M_{i}=-3.12+7.28(i-J)-0.65(i-J)^2\pm0.21\;\;\;\;\;\;\;\\
\nonumber
\end{eqnarray}

\noindent for the ranges 0.37 $\leq$ $i-z$ $\leq$ 1.84 and 1.92 $\leq$
$i-J$ $\leq$ 4.43 respectively.  These results agree within the
uncertainties with the relations derived by Williams et al. (2002).

\begin{figure}
\plotone{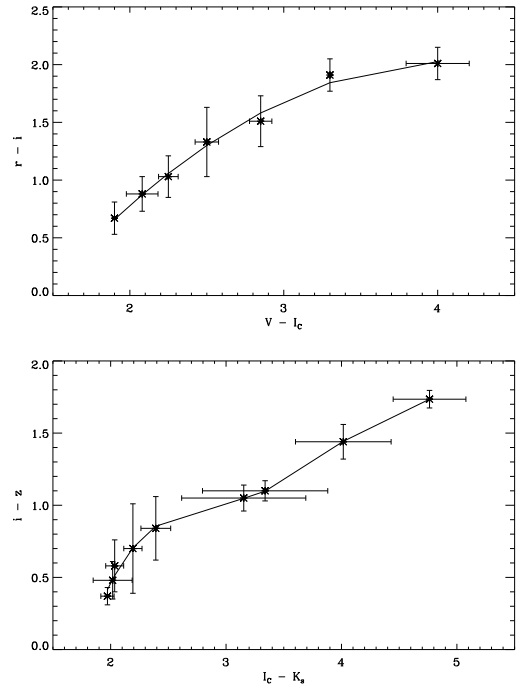}
\caption{Data and fits between SDSS and Johnson-Cousins/2MASS colors: $r - i$ 
versus $V-I_{\rm{C}}$ (upper) and $i - z$ versus $I_{\rm{C}}-K_{\rm{s}}$ (lower).  Stars are binned and averaged according to their spectral type.  For the $i - z$ vs.  $I_{\rm{C}}-K_{\rm{s}}$ relation, stars with spectral type later than M6 have been binned every 2 spectral classes (M7-M8, M9-L0).  Error bars represent the spread in color for each spectral type bin. Fits to both relations are given by Equations 3 and 4.}
\label{fig4}
\end{figure}

\subsection{Activity Relations}

Using the updated colors from Table \ref{tbl-1}, we follow the method of WHW and derive new relations for $r-i$ as a function of  $V-I_{\rm{C}}$ and $i-z$ as a function of  $I_{\rm{C}}-K_{\rm{s}}$.  As described in WHW, the $V$, $I_{\rm{C}}$ and $K_{\rm{s}}$ data come from nearby samples of M0-L0 stars (for complete sample details see WHW).  Individual stars have been binned according to their spectral types.  Very few stars with spectral classes later than M6 exist in these samples. Therefore, the late type stars have been binned every 2 spectral types (M7-M8 and M9-L0). Figure \ref{fig4} shows these relations together with our fits to
the color-color relationships.  The error bars represent the spread of the colors in each bin.  Note that the intrinsic spread of the population dominates the
scatter and is non-trivial. The fits can be described by:

\begin{eqnarray}
r  -  i &  = & -2.69 + 2.29(V-I_C) -0.28(V-I_C)^2\\
\nonumber\\
i - z & = & -20.6 + 26.0(I_C-K_s) - 11.7(I_C-K_s)^2\nonumber\\
 & & + 2.30(I_C-K_s)^3-0.17(I_C-K_s)^4\\
\nonumber
\end{eqnarray}

\noindent over the ranges 0.67 $\leq$ $r-i$ $\leq$ 2.01 and 0.37 $\leq$
$i-z$ $\leq$ 1.84 respectively.  The $i-z$ color derived from this relation has
an uncertainty in the fit of 0.10 magnitudes while the $r-i$ color has an 
uncertainty of 0.05 magnitudes.  The uncertainty of 0.05 magnitudes is 
small enough that photometric errors most likely dominate the error budget for the $r-i$ vs. $V-I_{\rm{C}}$ 
relation, rather than the uncertainty in the fit.

\begin{figure}[h]
\plotone{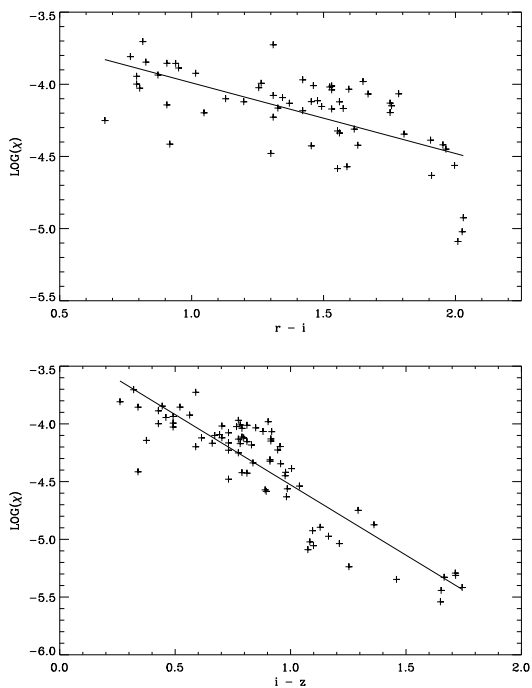}
\caption{$\log(\chi)$ versus $r - i$ (upper) and $i - z$ (lower) for stars in the WHW sample. Colors were transformed to the SDSS system using the color-color relations from Figure \ref{fig4} and Equations 3 and 4.  The $\log(\chi)$ fits are given in Equations 5 and 6.}
\label{fig5}
\end{figure}

Using the $\chi$ values and Johnson-Cousins/2MASS colors for each
star as given in WHW, we apply Equations 3 and 4 to transform the colors of the
WHW stars to the SDSS system. Fits to $\log(\chi)$ in $r-i$ and
$i-z$ are presented in Figure \ref{fig5} and described by the
equations:

\begin{eqnarray}
\log(\chi) & = & -3.50 - 0.49(r-i)\\
\log(\chi) & = & -3.31 - 1.22(i-z)\\    
\nonumber
\end{eqnarray}

\noindent over the ranges 0.67 $\leq$ $r-i$ $\leq$ 2.01 and 0.37 $\leq$
$i-z$ $\leq$ 1.84.  The uncertainty in the fit for $\log(\chi)$ is 0.22
for both the $r-i$ and $i-z$ colors.  We emphasize
that the $r-i$ relation \emph{should not} be used for spectral types later
than M6, as the reddest points in Figure \ref{fig5} show significant
systematic deviations from the mean relation.




\section{Summary}

Using SDSS DR3 photometry and the low mass star samples of W04, H02
and WHW, we calculate updated average colors for low-mass stars and
derive new photometric parallax relations.  We stress that the
intrinsic scatter in the colors of M and L dwarfs is large and any
distances determined from photometric parallaxes carry significant
uncertainties.  With our updated colors, we also provide new color
transformations between $r-i$ and $V-I_{\rm{C}}$ and between $i-z$ and
$I_{\rm{C}}-K_{\rm{s}}$.  Equations for calculating $\chi$ (see WHW) are
provided to aid in studying the magnetic activity in cool
stars.  Again we warn the reader to pay careful attention to the
propagation of uncertainties and the boundaries in color space over
which our relations are relevant when using these results. These data
will allow for more accurate studies of the distribution of low mass
stars in the Galaxy.  With the bountiful SDSS data, we will be able to
probe the structure of the Milky Way in new ways using some of its
most minute constituents.

\acknowledgements
We would like to thank Kevin Covey, John Bochanski and Nicole Silvestri 
for their helpful discussions in bringing this paper to fruition.  
AAW would like to acknowledge the support of Julianne Dalcanton 
during this study.  SLH acknowledges the support of the NSF through
grant AST 02-05875.
Funding for the SDSS has been provided by the 
Alfred P. Sloan Foundation, the Participating Institutions, the 
National Aeronautics and Space Administration, the National Science 
Foundation, the US Department of Energy, the Japanese Monbukagakusho, 
and the Max Planck Society.  The SDSS Web site is http://www.sdss.org. 
The SDSS is managed by the Astrophysical Research Consortium (ARC) for 
the Participating Institutions.  The Participating Institutions are 
University of Chicago, Fermilab, the Institute for Advanced Study, 
the Japan Participation Group, Johns Hopkins University, Los Alamos 
National Laboratory, the Max-Planck-Institut f\"{u}r Astronomie, the 
Max-Planck-Institut f\"{u}r Astrophysik, New Mexico State University, 
University of Pittsburgh, Princeton University, the US Naval Observatory, 
and the University of Washington.  This publication makes use of data products from the Two Micron All Sky Survey, which is a joint project of the University of Massachusetts and the Infrared Processing and Analysis Center/California Institute of Technology, funded by the National Aeronautics and Space Administration and the National Science Foundation.

\clearpage


\begin{thebibliography}{}
\bibitem[Abazajian et al.(2003)]{dr1} Abazajian, K, et al. 2003, \aj, 126, 2081
\bibitem[Abazajian et al.(2004)]{dr2} Abazajian, K, et al. 2004, \aj, 128, 502
\bibitem[Abazajian et al.(2005)]{dr3} Abazajian, K, et al. 2005, \aj, 129, 1755
\bibitem[Bochanski et al.(2005)]{boo05} Bochanski et al. 2005, in preparation
\bibitem[Covey et al.(2005)]{cov05} Covey et al. 2005, in preparation
\bibitem[Fukugita et al. (1996)]{fuk96} Fukugita, M., Ichikawa, T., Gunn, J.E., Doi, M., Shimasaku, K., \& Schneider, D.P. 1996, \aj, 111, 1748
\bibitem[Gunn et al. (1998)]{gun98} Gunn et al. 1998, AJ, 116, 3040
\bibitem[Hawley, S.L. et al.(2002)]{ha02} Hawley, S.L., et al. 2002, \aj, 123, 3409
\bibitem[Hogg et al. (2001)]{hogg01} Hogg, D.W., Schlegel, D.J., Finkbeiner, D.P., \& Gunn, J.E. 2001, \aj, 122, 2129
\bibitem[Juric et al.(2005)]{ju05} Juric, M. et al. 2005, in preparation
\bibitem[Pier et al. (2003)]{pier03} Pier, J.R., Munn, J.A., Hindsley, R.B., Hennessy, G.S., Kent, S.M., Lupton, R.H., \& Ivezi\'{c}, \v{Z}. 2003, \aj, 125, 1559 
\bibitem[Raymond et al. (2003)]{ray03} Raymond, S.N., et al. 2003, \aj, 125, 2621
\bibitem[Smith et al. 2002]{smi02} Smith, J.A., et al. 2002, \aj, 123, 2121
\bibitem[Stoughton et al. (2002)]{Stou02} Stoughton, C. et al.. 2002, \aj, 123, 485
\bibitem[Walkowicz, Hawley \& West (2004)]{whw04} Walkowicz, L.M, Hawley, S.L. \& West, A.A. 2004, \pasp, 116, 1105
\bibitem[West et al.(2004)]{we04} West, A.A., et al. 2004, \aj, 128, 426
\bibitem[West et al.(2005)]{we05} West. A.A., et al. 2005, in preparation
\bibitem[Williams et al. (2002)]{wil02} Williams, C.C. et al. 2002, BAAS, 34, 1292
\bibitem[York et al.(2000)]{yor00} York, D.G. et al. 2000, \aj, 120, 1579
\end{thebibliography}
\end{document}